\documentclass[manuscript, screen, nonacm]{acmart}

\AtBeginDocument{%
  }

\setcopyright{acmlicensed}
\copyrightyear{2018}
\acmYear{2018}
\acmDOI{XXXXXXX.XXXXXXX}

\acmConference[Conference acronym 'XX]{Make sure to enter the correct
  conference title from your rights confirmation emai}{June 03--05,
  2018}{Woodstock, NY}
\acmISBN{978-1-4503-XXXX-X/18/06}

\usepackage{caption}
\usepackage{subcaption}

\begin{document}

\title{Agency and Morality as part of Text Entry AI Assistant Personas}

\author{Andreas Komninos}
\email{akomninos@ceid.upatras.gr}
\orcid{1234-5678-9012}
\affiliation{%
  \institution{University of Patras}
  \city{Rio}
  \country{Greece}
}

\renewcommand{\shortauthors}{Komninos}

\begin{abstract}
This paper discusses the need to move away from an instrumental view of text composition AI assistants under direct control of the user, towards a more agentic approach that is based on a value rationale. Based on an analysis of moral dimensions of AI assistance in computer mediated communication, the paper proposes basic guidelines for designing the agent's persona.
\end{abstract}




\maketitle

\section{Introduction}
LLM technology has introduced a new stream of air in the sails of conversational user interfaces (CUIs), where humans engage in interactions with artificial agents, in order to accomplish tasks or goals. LLM-driven assistants are now interjecting CUI elements into traditional computer-mediated communication (CMC) interfaces, for example the instant messenger or SMS application, as message composition is now gradually affected by the integration and interaction with these assistants through the device keyboard, forming part of the text entry process. Let us call these text-composition assistants \textit{CHATs} (Clever Helper for Assisted Texting). Their integration with the keyboard allows us to quickly correct spelling and grammatical mistakes, or change the prose to a linguistic style more appropriate for the context of the conversation, or even generate entire responses without entering any text ourselves. While seemingly innocuous, CHATs raise significant ethical questions which should be addressed through careful design of both functional aspects of the integration, but also of the persona exhibited by the CHATs tasked with the modification or generation of text on behalf of the user. Consequently, I argue that design only superficially relates to the behaviour of the CHAT. In fact, design choices in the implementation of such tools, ultimately serve to design the persona of the user themselves.

\section{Five hundred million Cyrano de Bergeracs}
In the 1897 play "Cyrano de Bergerac", also adapted for cinema and inspiring numerous derivative works (perhaps most famously the comedy film \textit{Roxanne} starring Steve Martin), the tongue-tied but handsome Christian de Neuvillette is assisted by the intelligent and eloquent, but notoriously ugly Cyrano, to compose letters in order to romance and win over the beautiful Roxanne. The ploy is successful, but leads to a tragic outcome, as Cyrano is only much later revealed to be the author of the love letters. The revelation of Christian and Cyrano's deception and their ultimate deaths upend Roxanne's understanding of her relationship to the two men and her sentiments towards them, leaving her in a limbo of confusion and despair. Throughout the play, the concept of deception plays a central role, as a means used by characters to honourable ends (e.g. Cyrano), but also for personal advancement and gain, whether intentionally (as in the case of Christian) or inadvertently (in the case of Roxanne, who desires to idealise Christian) \cite{hauserMindingBehaviourDeception1997}.

We can find several parallels between this play and the 
abilities afforded by today's modern technology to assist text composition. Take for example the quality of machine translations, which can enhance one's ability to perform CMC over a foreign language with a native speaker, say for the purposes of booking an AirBnB and then arranging to meet the host at the location for the check-in \cite{carvalhoAttitudesMachineTranslation2023}. The deception that the user can communicate in the local's language is well-intended, as translation is used to support communication that might otherwise have been impossible. But it does create expectations in the local person, which only hold until the two persons meet face to face - then, expectations are shattered like those of Roxanne when she realises that Christian is nowhere near as eloquent in real life, as in the letters he sent her. In our AirBnB example, the native speaker promptly realises that the other party cannot, in fact, speak the language. Although communication in this way has, until recently, required use of disjointed tools (e.g., the messaging application, and a separate application in which text to be translated is copy-pasted), more recently, we have witnessed the integration of CHATs in OEM virtual keyboards, making them accessible inside any application. Galaxy AI was the first such example, allowing users to re-style original texts for prose before sending (e.g. professional, friendly) and automatically fix any spelling or grammatical mistakes in the text, or even to autogenerate entire replies to incoming messages. Google and Apple have also recently introduced similar features in mainstream devices. Like a modern Cyrano, these implementations offer "text-on-demand" which the modern Christian (the user) may choose or discard, without being any wiser as to what makes these texts more or less appropriate.

These concerns are not new. In the past, we have relied on third-party multilingual speakers to correct or compose our messages to others. We have asked for third-party advice on the tone and content of messages to be sent by us (e.g. friends or colleagues) \cite{careyExploringMentalModels2019}. We have frequently adopted internal alternative personas, which may not necessarily reflect our true sentiments or intentions, or enhance our social desirability while maintaining what we believe as authentic aspects our psyche \cite{bullinghamPresentationSelfOnline2013}. However, all these modes of "assistance" have required deliberation and reflection. Even the need to disclose messages to a (trusted) third party when asking for assistance, provided time and space for discourse and thought about how much to disclose, or what and how to write (e.g. taking “\textit{387 keystrokes to get to ‘Hey’}” \cite{rudderDataclysmWhoWe2014}. The automation and ready availability of LLM assistance removes these barriers, inhibitions and opportunities for reflections: It's right there, it's private to us, why not use it? While in the past it has been possible to bring a Cyrano into our lives, it is now easier, more convenient, more private and more tempting than ever to do so massively and at scale.

\section{Truth never damages a cause that is just.}
From a moral perspective, it's clear that ethical questions about the use of CHATs stem from the potential for deception, which in turn may be willful, or inadvertent \cite{fallisLyingDeceptionTheory2011}. Viewed from Kant's categorical imperative perspective, would we \textit{will} that the maxim "it should be permissible to use AI assistance during text composition" become a universal law? If everyone used CHATs with the intention of improving communication and supporting the user's authentic intent, we might see clearer and more effective CMCs, benefiting all of humanity and allowing the world to continue to function without logical contradictions. This would also be morally permissible from a utilitarian point of view. On the other hand, if CHATs were used to deceive, in a way that non-authentic perceptions of the communicating parties were created, then a violation of Kant's principle of never using people as a means, but only as an end unto themselves, is clearly present. Universal use of CHATs may also raise also questions of legal nature. To date, communications are admissible in court under the assumption that they are the product of autonomous will. However, if messages are generated automatically or altered by AI assistance, it is not clear where the boundary of responsibility may lie.

More importantly, there is a danger that universal use of CHATs, especially through models which are not trained or fine-tuned with individual users' authentic texts, may homogenise communication styles, and may lead to loss of linguistic diversity and personal expression, particularly when complete replies are auto-generated, rather than re-worded \cite{deroockBecomeObjectObjects2024}. This would diminish human agency, autonomy and individuality, therefore completely undermining human ability to perform duties derived from our moral values. This deficit stems from the lack of reflection and effort put into communication. For example, under Uncertainty Reduction Theory, the language of messages shapes our model of the social environment, hence delegation of linguistic expression to AI limits our ability to form models of the other parties, and hence a meaningful relationship to them \cite{baxterEngagingTheoriesInterpersonal2008}.

I would hazard to express a concern that beyond these dangers, there is a more fundamental issue at stake, namely the erosion of our ability to form moral values in the first place. If all our communication flaws are smoothed out by CHATs, we lose the opportunity to experience the friction of imperfect discourse and its unpleasant impacts, which can force us to reflect and therefore empower our ability to consider our own values, the way we articulate them, and ultimately fulfill our capacity for independent critical and ethical thinking \cite{narvaezChapterMoralIdentity2009}. Since in the modern age so much of our lived experience is constituted through CMCs, the CHAT becomes an indispensable crutch, without which we are left helpless both in online settings, but most importantly in the real world. A diminished capacity to independently articulate our thoughts, may lead to an inability to reveal our values to ourselves, and thus engage in any empathic and meaningful conversation with others, even before losing the ability to form models about them. 

\begin{figure}
     \centering
     \begin{subfigure}[b]{0.45\textwidth}
         \centering
         \includegraphics[width=\textwidth]{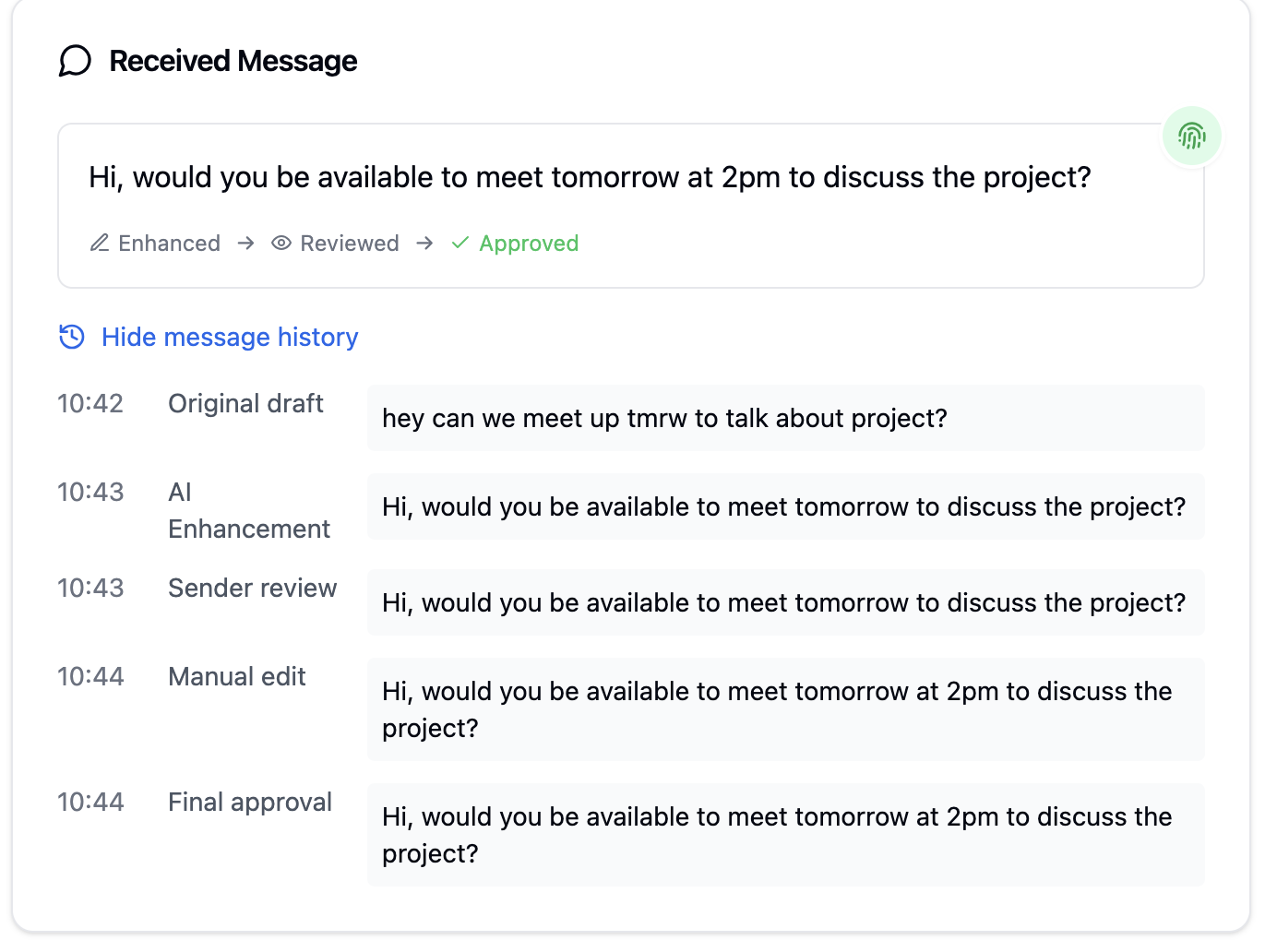}
         \caption{View of a received message with CHAT assistance}
         \label{fig:disclosure}
     \end{subfigure}
     \hfill
     \begin{subfigure}[b]{0.45\textwidth}
         \centering
         \includegraphics[width=\textwidth]{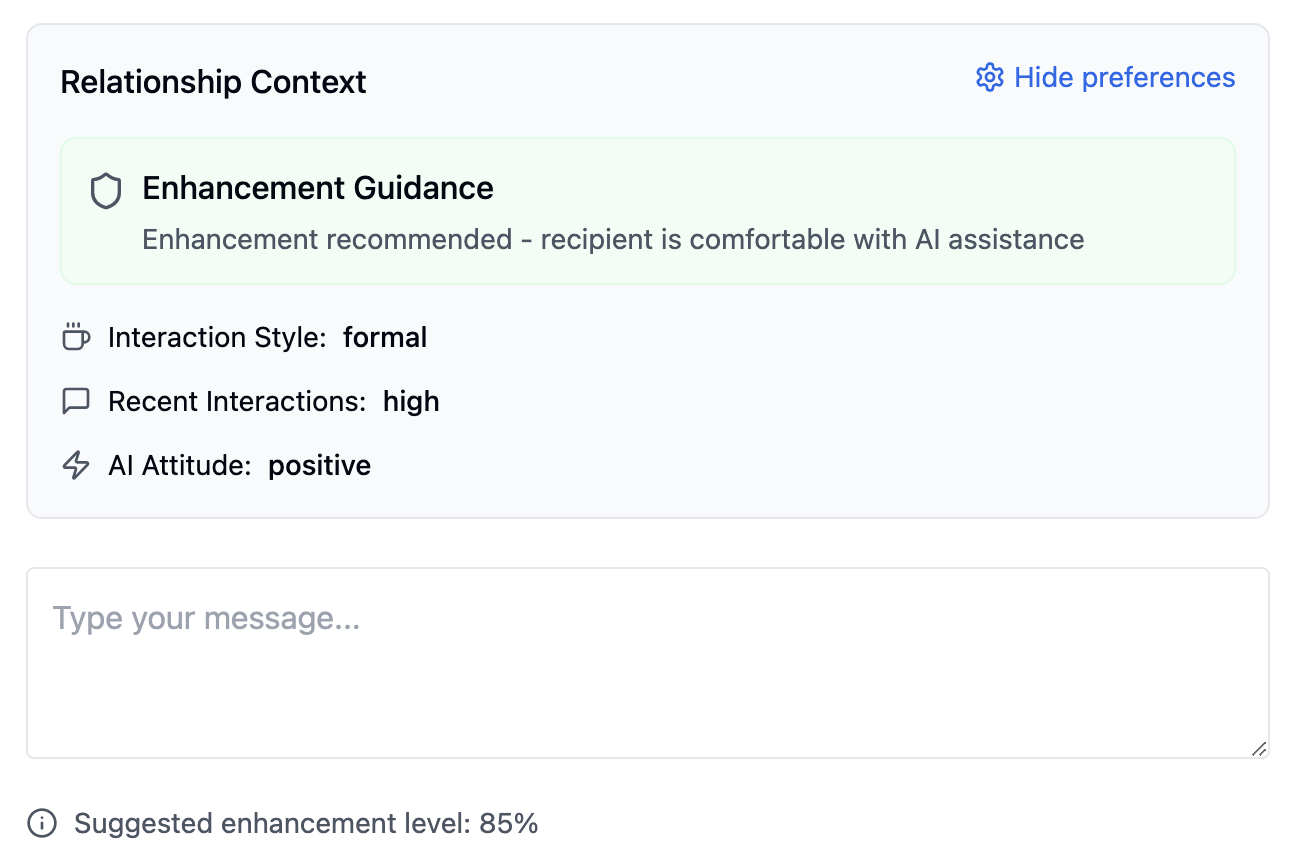}
         \caption{CHAT supporting the user to make informed and respectful decisions}
         \label{fig:guidance}
     \end{subfigure}
        \caption{Examples of disclosure behaviour and preference negotiation - sharing.}
        \label{fig:design}
\end{figure}

\section{Persona characteristics of a CHAT}
So far we outlined the opportunities, but also threats arising from CHATs. Let us now turn to some thoughts about how design may mitigate some of these problems. Firstly, if we are to apply in our technology the Kantian principle of treating people as an end unto themselves, then CHATs must bear responsibility towards both communicating parties, and not just the party employing their services to improve their text. Otherwise, the potential for misusing assistants to deceive, is significant. As such, I argue that we should depart from the implementation of CHATs under \textit{instrumental} rationality (e.g. the mere "use" of a Cyrano as a means to achieve an end), and adopt a \textit{value} based rationality, in which CHATs are to be treated as an independent third party in the CMC, with agency in the role of a mediator. Of course, to be effective, such agents should adopt a carefully designed persona with distinct tone, voice, and personality. 

The first characteristic of this personality must be transparency, which is a key constituent for managing expectations and for maintaining integrity and authenticity in CMCs so as not to fall in the trap of deceptive use. A CHAT is a mediating party in the communication process between two persons, therefore its presence, capabilities and actions should be clear to both sender and receiver. Willful readiness to reveal this information to both parties, should be part of the agent's persona. In Fig. \ref{fig:disclosure}, it's not just the message that is transmitted, but also metadata showing that the message was co-authored with a CHAT, reviewed and approved by the sender. If the receiver wishes, they can query the sender's CHAT to obtain more information.
Second, the mediator's presence and involvement must be consensual through a negotiation process from both parties. A receiver should be informed of a sender's intention to employ (or that they are already making use of) assistance, and should consent or opt out to it, either on a per-conversation basis, or as a general preference. An agent's persona should be designed in a way to handle this negotiation on behalf of the agent's user, in a respectful and non-coercive manner. 
Third, the agent should be supportive rather than prescriptive. Further from simply providing suggestions, explanations of the rationale to the user must also be offered, therefore allowing them to make informed choices about the text recommended to them, and to learn in the process. In Fig. \ref{fig:guidance} the user of a CHAT is shown how the CHAT can support them, respecting the preferences of the receiver and other contextual information that the CHAT has been able to obtain (e.g. via reading the message history). 
Fourth, the mediator bears responsibility to reinforce and promote ethical and critical thinking for the party employing it, mitigating the negative effects of the continuous "smoothing out" of communication problems. This personality aspect includes elements of self-restraint, making the agent non-cooperative in some cases. For example, the agent may refuse to intervene in the case of communicating with a new contact, ceding to human touch as the start of a conversation. It may gradually offer progressively advanced enhancements, from simple grammar to linguistic styles, as the agent learns more about the contact through communication history, and perhaps by explicitly asking the user themselves or querying the receiver's preferences and attitudes towards use of CHATs. In these examples I have attempted to lay out the basic characteristics of the persona. How these persona traits are shown to the user, remains to be further designed, whether as a conversational inteface (e.g. akin to how we might ask a friend's advice on how to compose, engaging in a feedback loop), or whether as a range of options or commands that users may be able to apply (or not) under different contexts.

\section{Conclusion}
Current implementations of CHATs reflect an instrumental view of the system and the humans affected by it. By offering these abilities at hand as simple commands, the assistant exhibits agency without autonomy, steering the user's own persona towards that instrumentality, and leading to the gradual dehumanisation of the communication process, and of the users. Thoughtful design might mitigate some, if not all emergent dangers, though technical implementation of any design will be hard. In the name of progress, technical abilities are introduced expediently in core aspects of societal functions. I wonder whether such "progress" drives individuals towards achieving the ideal Heideggerian state of \textit{Dasein}, or pushes them ever closer to a state of inauthentic, amorphous and depersonalised existence as the\textit{Das-Man}.

\section{Publication details}
This paper was presented at Designing AI Personalities: Enhancing Human-Agent Interaction Through Thoughtful Persona Design, Workshop in conjuction with ACM Mobile \& Ubiquitous Multimedia, Stockholm, Sweden, Dec. 1, 2024 

\bibliographystyle{ACM-Reference-Format}
\bibliography{bibliography}

\end{document}